\documentclass[11pt,a4paper,english]{article}
\usepackage{url}
\usepackage{natbib}
\usepackage{amssymb}
\usepackage{graphicx}
\usepackage{color}
\usepackage{multirow}
\usepackage{babel}
\usepackage{authblk}
\usepackage{blindtext}
\usepackage{float}
\restylefloat{table}

\title{Spatial Point Pattern Analysis of the Unidentified Aerial Phenomena in France}
\author[1]{Thibault Laurent\thanks{Thibault.Laurent@tse-fr.eu}}
\author[2]{Christine Thomas-Agnan\thanks{Christine.Thomas@tse-fr.eu}}
\author[3]{~ \\Micha\"{e}l Vaillant\thanks{Michael.Vaillant@gmail.com}}

\affil[1]{Toulouse School of Economics (GREMAQ/CNRS)}
\affil[2]{Toulouse School of Economics (GREMAQ)}
\affil[3]{Meta-Connexions}

\begin{document}
\maketitle
\begin{abstract}
We model the unidentified aerial phenomena observed in France during the last 60 years as a spatial point pattern. We use some public information such as population density, rate of moisture or presence of airports to model the intensity of the unidentified aerial phenomena. Spatial exploratory data analysis is a first approach to appreciate the link between the intensity of the unidentified aerial phenomena and the covariates. We then fit an inhomogeneous spatial Poisson process model with covariates. We find that the significant variables are the population density, the presence of the factories with a nuclear risk and contaminated land, and the rate of moisture. The analysis of the residuals shows that some parts of France (the Belgian border, the tip of Britany, some parts in the South-East, the Picardie and Haute-Normandie regions, the Loiret and Corr\`eze departments) present a high value of local intensity which are not explained by our model.
\end{abstract}
\section{Introduction}
An Unidentified Aerospace Phenomenon  (UAP) (in French \textquotedblleft Ph\'enom\`ene A\'erospatial Non identifi\'e\textquotedblright ),
correspond to a phenomena that does not currently find legitimate explanations, most often due to a lack of information, but also,
in rarer cases, due to our actual limitations in terms of scientific knowledge.

In France, once a person has been a witness to a UAP, she or he has the possibility to report at the Gendarmerie The witness is asked to fill
in a detailed survey to provide information such as date, time, place, duration, orientation, shape, size, trajectory, witness' distance to the phenomenon, etc.
The investigation is then handed over to the GEIPAN (\url{http://www.geipan.fr/}), a unit of the French Space Agency CNES, whose main mission is to validate the
information provided by the witness and to determine the nature of the UAP. In addition, the GEIPAN classifies each UAP into 4 categories A/B/C/D,
forming a kind of scale, which goes from perfectly known and determined (A) to unknown and undetermined (D), after investigation.

It should be noted that even today, 19.5\% of UAPs remain undetermined after investigation which is frustrating of course for both the witness and the scientist.

\subsection{A new strategy to constrain the space of hypothesis}
After over 50 years of lack of progress in the field of Unidentified Aerospace Phenomena,
we decided to test new ways of analysis, as well as an innovative approach, based on a global analysis,
so as not to be dependent on isolated testimonies. This approach stems from a simple observation: Aerospace Phenomena,
whatever they are, are only the products of two origins:
\begin{itemize}
\item a. endogenous phenomena, created within the observed environment,
when favorable local conditions allow the emergence of a \textquotedblleft rare\textquotedblright{}
phenomenon (the level of scarcity is subjective to the observer)
\item b. exogenous phenomena, created outside of the observed environment.
Which can in turn be divided into two categories:

\begin{itemize}
\item b1. phenomena which come in the environment because of local conditions
(a local attractor). This is especially relevant if this phenomenon
(which can be regarded as a complex system) remains durably in relation
with the environment in which it is observed.
\item b2. phenomena which cross the environment in a \textquotedblleft forced\textquotedblright{}
way. An interaction with the environment is an unintended consequence.
\end{itemize}
\end{itemize}

\begin{figure}[h!]
\centering
\includegraphics[scale=1.1]{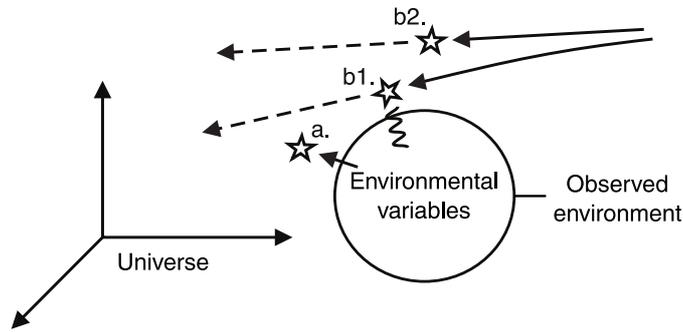}
\caption{\label{phenomena-sources} Endogenous and exogenous UAP in relation
with the observed environment.}
\end{figure}

Since environmental variables (which cover environmental resources)
are likely to be involved in the explanation of many phenomena, our
strategy has been to seek environmental variables that could correlate
with the presence of UAPs. In this way, we want to reduce the space
of possible explanations for Unidentified Aerospace Phenomena and
also propose new areas to think about.

\subsection{Selecting variables}
This bring us to consider two categories of variables:
\begin{itemize}
\item Variables which are involved in environmental characterization, including,
on the one hand, those of anthropogenic nature, signing human activity, and on the other,
those of biological nature, signing natural ecosystems.
\item Variables of systemic nature, that could significantly and permanently
alter the environment.
\end{itemize}

The table \ref{covlist}~summarizes the covariates we have selected for
our analysis:

\begin{table}[h!]
\caption{\label{covlist} {*} Please note that we are in the range of risks
conventionally named \textquotedblleft Nuclear, Biological, and Chemical\textquotedblright{}
(NBC) {*}{*} Within the natural hazard range we did not test: seismic
risks, floods and hydrogeological risks, volcanic hazards and forest
fires. These could be added in future studies. }
\centering
\begin{tabular}{l|rr}
  \hline
  Covariates &  environmental characterization &  systemic risks\\
  \hline
  Anthropogenic  & Density of population  &  Number of nuclear sites{*} \\
                           & Number of airports     &  Number of polluted sites{*} \\ \hline
                           & Percentage of wetlands &                               \\
Environmental   & Percentage of forests  & N/A{*}{*}          \\
                           &  Sunshine              & \\ \hline 
\end{tabular}
\end{table}

Afterward, we also decided to use UAP As (identified phenomena) as a covariate.
Our hypothesis is that, if, in some areas, the number of testimonies relies on
socio-psychological characteristics (eg. some people will talk easily while others dare not talk),
then the variation of testimonies should also be reflected in the number of registered UAPs,
and therefore we should observe a certain level of correlation between UAPs A/B/C/D.
In that sense, using UAP As as a covariate may help us determine whether
an abnormal variation of UAP Ds is explained by socio-psychological resistances or
on the contrary by socio-psychological openness.

\subsection{Geographical and historical scope of our study}

An UAP may have been observed by several persons. In that case, the
localization of the phenomenon is given by the centroid of the convex
hull of the points of observation. If the greatest distance of the
centroid from the vertices is morer than 20 kilometers, the UAP will
be excluded from the study because of the lack of accuracy of the
coordinates (representing less than 8.1\% of the cases
of our database). From 1951 to 2013, 1969 UAPs have
been observed in metropolitan France (most of the data are available
at \url{http://www.cnes-geipan.fr/fileadmin/documents/Export_etudedecas.csv}).

\begin{figure}
\centering
\includegraphics[trim=0cm 0cm 0cm 0cm, clip=true]{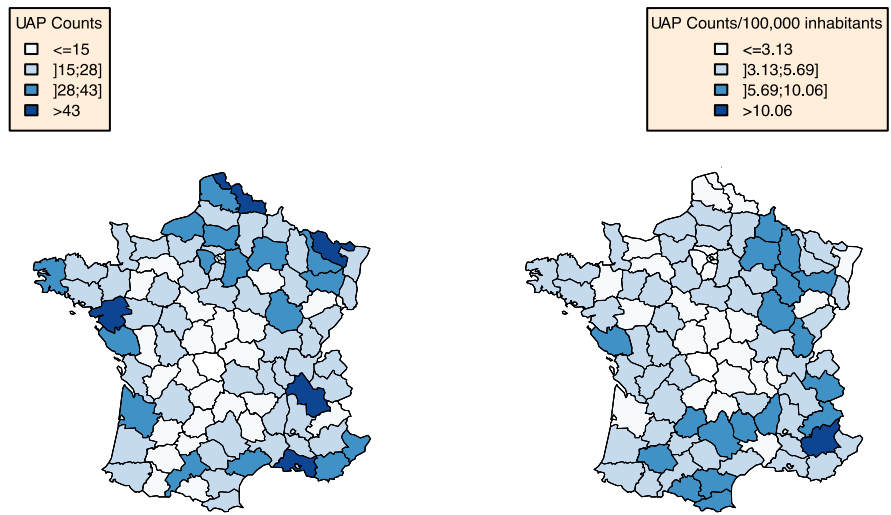}
\caption{\label{carto1} Locations of the 1969 UAPs observed in France from 1951 to 2013 (source: GEIPAN and INSEE).}
\end{figure}

Our study does not include those UAPs observed in Corsica or over the Ocean
 or the Mediterranean sea so as to facilitate the analysis of the spatial point pattern in a connex window.
 All 1969 UAPs have been represented by department in the left panel of  Fig.~\ref{carto1}.
 The administrative boundary has been downloaded at \url{http://www.gadm.org/}. We first
note that the intensity of the UAP is not homogeneous in space.
In our case, the departments of the North tend to report at large
number of UAPs, and so do the departments whose main city is a metropolitan
area (Bouches-du Rh\^one, Gironde, Is\`ere). The variance of the number
of inhabitants per department is large and for that reason, we represent
in the right panel of Fig.~\ref{carto1} the UAP counts for
100~000 inhabitants (We use the population count for 1990. The official
statistics about the population in France are given by the French
national statistical institute INSEE at \url{http://www.insee.fr/fr/themes/detail.asp?reg_id=99&ref_id=estim-pop}.).
The right hand side map is not uniformly colored which indicates
that the density of the population does not alone explain the intensity of the UAPs.

The main idea of this paper is to find out whether some
of the above-mentioned covariates can explain the intensity of the UAPs.
However, we are not considering all the categories of UAPs but only those UAPs whose nature has not been determined:
they are internationally known as UFOs  (see Section~\ref{section2}).
The data used to construct the covariates are provided by different national French statistics services.
Extensive work was conducted for aggregating the data, and is described in Section~\ref{section3}.
Section~\ref{section4} will present the main results of the modeling.
For this study, we used the statistical software R \citep{Rcore}
and mainly functions from the \texttt{spatstat} package \citep{Baddeley}.

\section{Descriptive statistics of the unidentified aerial phenomena}\label{section2}
\subsection{Definition and nature of unidentified aerial phenomena}
\begin{figure}
\centering
\makebox{\includegraphics{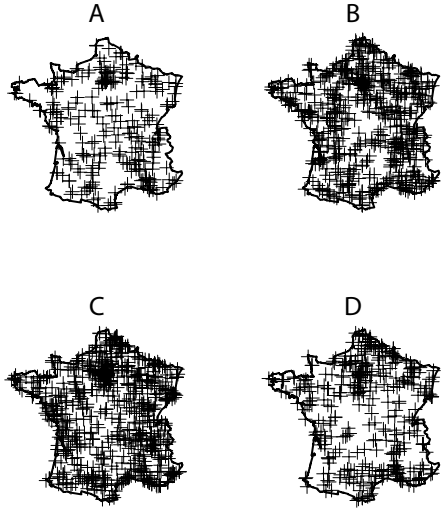}}
\caption{\label{allPAN} Locations of the UAPs according to the GEIPAN classification. Each UAP is located at the centroid of the commune where the UAP was observed.}
\end{figure}

The classification of the UAPs made by the GEIPAN investigators as represented in Fig.~\ref{allPAN} is as follows:
\begin{itemize}
\item UAP As: the case has been explained unambiguously (237 observations - $12\%$ of the sample),
\item UAP Bs: the case is probably identified (581 observations - $29.5\%$ of the sample),
\item UAP Cs: the observation is non-identifiable because of the lack of data (770 observations - $39\%$ of the sample),
\item UAP Ds: the observation is non-identifiable (381 observations - $19.5\%$ of the sample).
\end{itemize}
The reasons for classifying a phenomenon as a UAP A or a UAP B can be: observation of a star, observation of the passage of Thai lanterns,
observation of a re-entry, observation of an airplane during landing, etc.
UAP Cs essentially correspond to the phenomena which have been observed by only a few witnesses ($80\%$ of UAP C cases have been observed by only one or two witnesses).
The UAP Ds have been subdivided into two categories (see \url{http://www.geipan.fr/index.php?id=201}), but we will not take this into account in this study.
The distribution of the UAPs has evolved over the years because the information available for older cases is rarely as complete as for recent cases.
Thus, Table~\ref{time} shows that during the last 10 years, the proportion of UAPs classified as UAP Ds has substantially decreased.
Moreover, the analysis of a $\chi^2$ test indicates that the distribution of the UAPs depends on the time period (The value of the $\chi^2$ is equal to 233, the null hypothesis of independence between the time and the classification is rejected with a p-value lower than 0.05).

\begin{table}
\caption{\label{time}Conditional distribution of the UAP D depending on the period of observation. The table has been generated with the cases for which the date of the phenomenon was available.}
\centering
\small{
\begin{tabular}{rrrrrrrr}
  \hline
 Period & 50-60 & 60-70 & 70-80 & 80-90 & 90-00 & 00-10 & 10-20\\
  \hline
  UAP D / Total UAP & 11/34 & 9/32 & 149/691 & 105/422 & 80/325 & 23/306 & 4/159  \\
  Proportion & 0.32 & 0.28 & 0.22 & 0.25 & 0.25 & 0.08 & 0.03  \\  \hline
\end{tabular}
}
\end{table}

The complete descriptions of the UAPs are given on the GEIPAN website.  Some blogs (see for example \url{http://sceptic-ovni.forumactif.com/forum}) propose alternative explanations (often of astronomical nature) for the UAPs. However, it is not the purpose of this paper to explain the nature of the UAPs. We will instead focus on the spatial point pattern analysis of the UAP Ds and the modeling of the intensity of the UAP Ds by some environmental covariates. The next section presents the kernel smoothed intensity of the UAP Ds.

\subsection{Kernel smoothed intensity of the UAP Ds}
We consider the 381 observed UAP Ds as a realization of a spatial point pattern $\{x_1,\ldots,x_n\}$. The observation window $A$ used corresponds to the polygon of metropolitan France whose area is equal to $540~461$ $km^2$. We use a Lambert Conic Conformal projection with two standard parallels to facilitate the computation of distances in kilometers. Thus, the average intensity is equal to 0.000705 cases per square kilometer. Complete Spatial Randomness (CSR) describes a point process whereby point events occur within a given study area in a completely random fashion \citep{Diggle}. Fig.~\ref{allPAN} suggests that this spatial point pattern does not follow the CSR assumption: A $\chi^2$ test of CSR using a $6 \times 6$ quadrat counts test gives a value of $\chi^2$ equal to 198 with a p-value lower than $0.05$. We also computed simulation envelopes for the Ripley's $K$ function \citep{ripley:1981}: the empirical Ripley's $K$ function was not included in the envelope obtained with 100 simulations.

\begin{figure}
\centering
\makebox{\includegraphics[trim=0cm 1.5cm 0cm 0cm, clip=true]{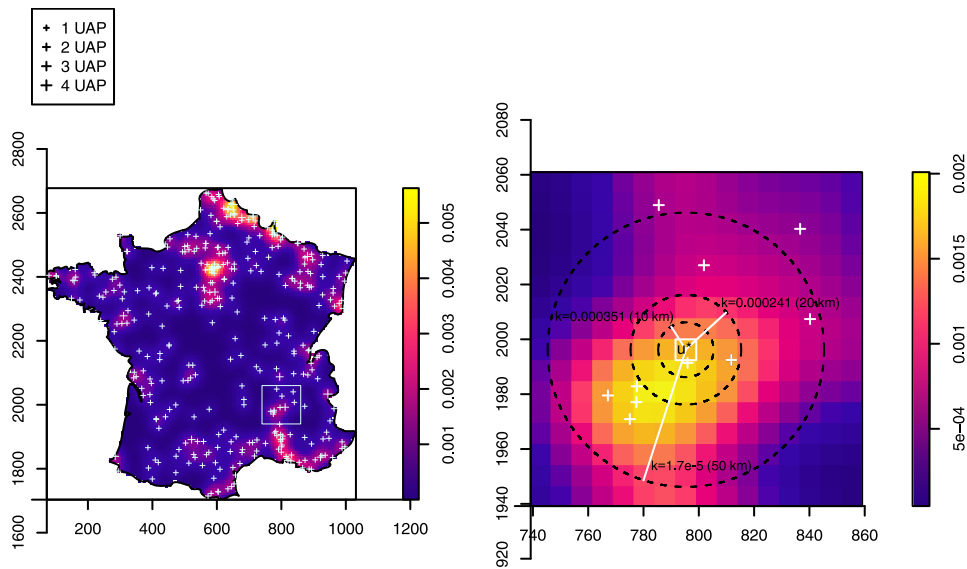}}
\caption{\label{Fig2} Non-parametric estimation of a spatially varying intensity for the UAP Ds with a value of $\sigma=20$. The right panel shows an example of computation of $\lambda^{*}$ at $u^{*}$.}
\end{figure}

We represent on the left panel of Fig.~\ref{Fig2}, the non-parametric estimation of a spatially varying intensity as defined by \cite{Diggle}. Thus the intensity value at a point $u$ is estimated by:
\begin{center}
$\lambda^{*}(u) = \displaystyle\sum_{i=1}^n e(x_i)k(x_i - u)$
\end{center}
where $k$ is the Gaussian smoothing kernel and $e(x_i)$ is an edge correction factor. The function \textit{density.ppp()} of \textit{spatstat} computes by default the intensity $\lambda^{*}$ on a square window of $128 \times 128=16384$ pixels. In our case, the number of pixels used for computation is equal to 9480 (\emph{i.e.} 16384 pixels minus the number of pixels not in $A$).  The size of a pixel is $7.49 \times 7.61$ kilometers. To select the smoothing bandwidth for the kernel $k$, we used indications given by the functions \textit{bw.diggle()} and \textit{bw.scott()}. The first function which minimizes the mean-square error criterion defined by \citet{Diggle1985}, recommends a value of $\sigma$ equal to 18 kilometers. The second function which uses Scott's rule for bandwidth selection for kernel density \citep{scott} gives a value of $\sigma$ equal to 77 kilometers in the $x$ direction and 100 kilometers in the $y$ direction. Finally, we choose a value of $\sigma$ equal to 20 kilometers. Let $u^{*}$ denote the pixel located in the rectangle $[791.58, 799.07] \times [1992.4, 2000]$ kilometers that we will use to illustrate the statistical methods presented in this paper. Fig.~\ref{Fig2} on the right panel represents an example of computation of $\lambda^{*}$ at pixel $u^{*}$.  We show that the contribution (ignoring edge corrections) to $\lambda^{*}$ of a spatial point $x_i$ 10 kilometers away from $u^{*}$ is equal to $k(x_i-u)= 0.000241$ whereas the contribution of an $x_i$ 50 kilometers away from $u^{*}$ is only equal to $0.00001748$. Finally, a value of $\lambda^{*}$ equal to 0.00255 at the pixel $u^{*}$ means that the expected number of UAPs at $u^{*}$ is equal to $7.49 \times 7.61 \times 0.00251=0.1453$. It is also important to mention that we recover the total number of cases by integrating the estimated intensity:
\begin{center}
$7.49 \times 7.61 \times \displaystyle \sum_{u=1}^{9480}\lambda^{*}(u)=378$
\end{center}

The analysis of the map reveals several local areas with a high concentration of the UAPs: first, around Paris and just above (Picardie and Haute-Normandie regions), we observe two clusters of observations. The Belgian border also seems to be more exposed as the tip of Brittany. There seem to be some other clusters in the South of France, it seems that along the Rh\^one river and in the Massif Central, there are also some clusters of UAPs. The next section presents the covariates which we consider to include for modeling the intensity of the UAPs.

\section{Description of the covariates}\label{section3}
All the covariates used in this study originate from several official statistical offices in France. The geographic coordinate systems were not necessarily the same. The different data are given with a different level of spatial resolution: communes, spatial point pattern, pixel, etc. For that reason, we will present in detail the data and the methods used to transform the data into pixel images which is the best way to make the data compatible.
\subsection{Population density}
\begin{figure}
\centering
\makebox{\includegraphics[trim=0cm 1.8cm 0cm 2cm, clip=true]{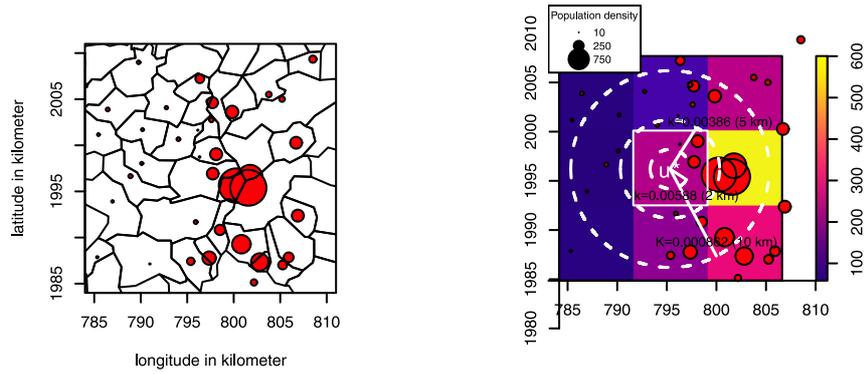}}
\caption{\label{Fig3} Spatial smoothing of the population density observed at the centroids of the communes. The right panel illustrates the values obtained at $u^{*}$ by the function $k$ with a value of $\sigma=5$ for different distances.}
\end{figure}

It seems clear that if the intensity of the UAP is so strong in the metropolitan area of Paris, this is partly due to a high value of population density in that area. This information is provided by the INSEE (see \url{http://www.insee.fr/fr/bases-de-donnees/default.asp?page=statistiques-locales.htm}). It is given in number of inhabitants per square kilometer for each of the 36~208 communes of metropolitan France. The reference date of the census corresponds to 1990. Thus, we have to assume that the population density did not significantly vary over the last 60 years. \citet{insee} takes the Pays de la Loire region as an example and shows that this hypothesis is not necessarily realistic; however, the spatial distribution of the population density is broadly stable. We note $\{v_1,\ldots,v_n\}$ the population density at locations $\{x_1,\ldots,x_n\}$ where the locations are the centroids of the communes (see \url{http://www.ign.fr/}).  The smoothed value at a location $u$ is (ignoring edge corrections):
\begin{center}
$g(u) = \displaystyle \frac{\sum_{i=1}^n k(u-x_i) v_i}{\sum_{i=1}^n k(u-x_i)}$
\end{center}
where $k$ is a Gaussian smoothing kernel, known as the Nadaraya-Watson smoother \citep{Watson1964-SRA,GVK025319159,Nada:1964}. The function used to compute this estimator is \textit{Smooth.ppp()}. Fig.~\ref{Fig3} illustrates the use of this function. We only represent the communes of France belonging to the given rectangle: $[785, 810] \times [1985, 2010]$ kilometers. The population density is represented by circles located at the centroid of the communes, with a size proportional to the values $v_i$ \citep{Tanimura}. On the right panel, we represent the Nadaraya-Watson smoother with a value of $\sigma$ equal to 5. The value of $\sigma$ returned by the function \textit{bw.smoothppp()}, which uses a least-squares cross-validation to select a smoothing bandwidth, is 2. With a value of $\sigma$ equal to 2 the map was not smooth enough, so we use $\sigma=5$. We also represent for different distances, the values of the function $k$ obtained at $u^{*}$ with a $\sigma$ equal to 5.

We represent the smoothed map on Fig.~\ref{covariates1}. The smoothed value $g$ has been computed for the 9480 pixels included in the window $A$. The discretization used for representing $g$ corresponds to the one commonly used at INSEE. It seems that the areas with a high population density are those which contain high numbers of UAP Ds. This observation can be confirmed by Table~\ref{densitytable} which represents the percentage of observed UAP Ds as a function of population density. $51.5\%$ of the UAP Ds are in zones of population density higher than 100 inhabitants per square kilometers whereas their area represents only $21.8\%$ of the French territory. Finally, a Kolmogorov-Smirnov test of goodness-of-fit of a Poisson point process model \citep{citeulike:390763} has been used with the function \textit{kstest.ppm()}. The value of the $D$ statistic is equal to 0.3023 with a p-value strictly lower than 0.05, showing that the covariate population density is significant to explain the intensity of the UAPs.
\begin{table}
\caption{\label{densitytable}Conditional distribution of the UAP Ds as a function of population density. Only 373 on 381 UAP Ds have been allocated to a class of population density because the function \textit{Smooth()} does not systematically give an estimation of the density on the border.}
\centering
\small{
\begin{tabular}{rrrrrr}
  \hline
 Inhabitants/$km^2$ & $<$=50 & ]50;100] & ]100;500] & ]500;2000] & ]2000;11224] \\
  \hline
  Percent of UAP Ds (out of 373) &   29.5 & 19.0 & 35.1 & 12.3 & 4.1 \\
Percent of Pixels (out of 9480) & 54.2 & 24.0 & 19.1 & 2.3 & 0.4 \\
   \hline
\end{tabular}
}
\end{table}

\subsection{Nuclear sites}
\begin{figure}
\centering
\makebox{\includegraphics[trim=0cm 0cm 0cm 0cm, clip=true]{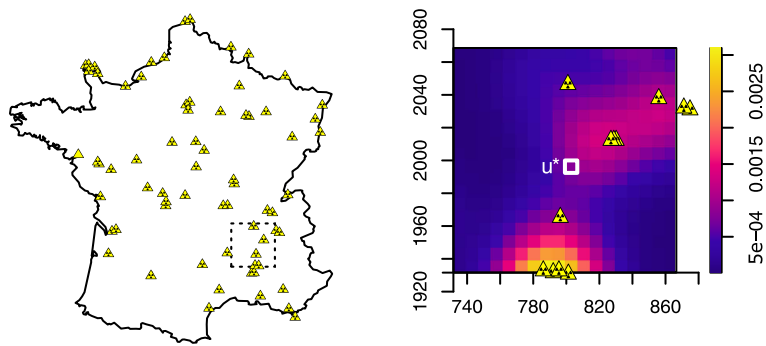}}
\caption{\label{nuclear} Locations of the 102 sites which present a nuclear risk in France. The right panel shows an example of computation for attributing to $u^{*}$ the kernel smoothed intensity with a value of $\sigma$ equal to 20 kilometers.}
\end{figure}

Fig.~\ref{nuclear} on the left panel represents the localization of the sites which present a nuclear risk in France. The data comes from the website \url{http://www.sortirdunucleaire.org/carte/}. There are 102 sites which present characteristics such as: nuclear power plants, uranium mining residues, storage and disposal of radioactive waste, etc. Firstly, we compute for each of the 9480 pixels included in the window $A$, the number of nuclear sites included in a neighborhood of 20 kilometers of the pixel. For this, we use the function \emph{dnearneigh()} of the package \emph{spdep} \citep{bivand}. Table~\ref{nucleartable} indicates that $12.3\%$ of the 9480 pixels have at least one nuclear site in their neighborhood. This percentage is equal to $21.2\%$ for the pixels which contain UAP Ds. Thus it clearly indicates that this covariate is potentially linked to the intensity of the UAP Ds.
\begin{table}
\caption{\label{nucleartable}Conditional distribution of the UAP Ds as a function of the number of nuclear sites within a 20-kilometer perimeter.}
\centering
\begin{tabular}{rrrrr}
  \hline
 Nuclear sites & 0 & ]1;5] & ]5;10] & ]10;13] \\
  \hline
 Percent of UAP Ds (out of 373)& 78.8 & 19.0 & 1.6 & 0.6\\
Percent of pixels (out of 9480) & 87.7 & 11.9 & 0.3 & 0.1\\
   \hline
\end{tabular}\end{table}

The idea for creating a pixel image of this data is to compute the kernel smoothed intensity for each of the 9480 pixels included in window $A$. The value of $\sigma$ has been taken equal to 20 kilometers as for the Kernel smoothed intensity of the UAP Ds. Fig.~\ref{nuclear} on the right panel shows an example for computing this covariate at $u^{*}$. The spatial Kolmogorov-Smirnov test of CSR rejects the hypothesis of CSR with a $D$ test statistic equal to 0.1299 and a p-value lower than 0.05, showing that the nuclear sites covariate is significant to explain the intensity of the UAPs individually.

Fig.~\ref{covariates1} represents the complete pixel image of the presence of nuclear sites in the neighborhood in region $A$.

\subsection{Contaminated land}
\begin{figure}
\centering
\makebox{\includegraphics[trim=0cm 0cm 0cm 0cm, clip=true]{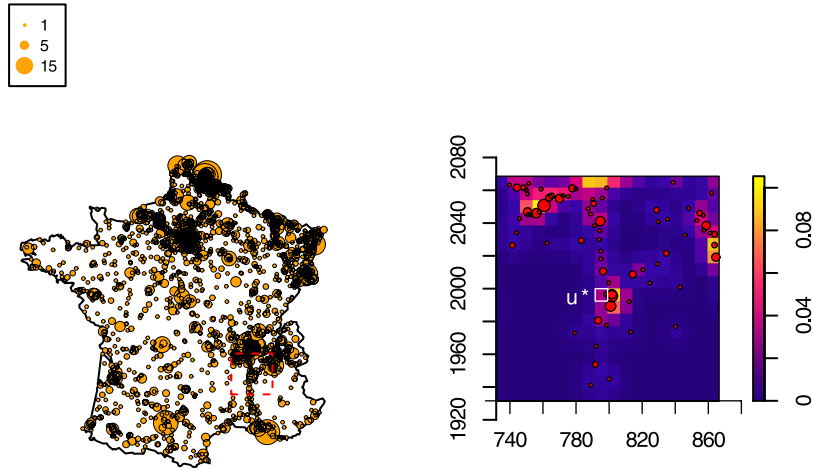}}
\caption{\label{contaminated} Locations of the 3~985 contaminated sites in France. The right panel shows an example of computation for attributing to $u^{*}$ the kernel smoothed intensity.}
\end{figure}

A contaminated site is a site with a perennial or potential risk to human health or the environment, due to pollution from a former or current activity. The origin of local pollution is usually industrial. At the end of 2007, there were 3~985 contaminated sites included in the window $A$, for which the French government has undertaken corrective action. The French department of statistics of the Ministry of ecology, sustainable development, transportation and housing is responsible of the publication of this data (see \url{http://basol.environnement.gouv.fr/}). The data are given as the number of contaminated sites per commune (see Fig.~\ref{contaminated}). Firstly, we will consider for each pixel $u$ included in window $A$, the number of contaminated sites in a neighborhood of 5 kilometers.  Table~\ref{contaminatedtable} indicates that $49.3\%$ of the UAP Ds are included in pixels where the number of contaminated sites is larger than 1 whereas this percentage is equal to $19.7\%$ for all the pixels included in window $A$. Thus, this covariate is potentially linked to the intensity of the UAP Ds.
\begin{table}
\caption{\label{contaminatedtable}Conditional distribution of the UAP Ds as a function of the number of contaminated sites within the perimeter.}
\centering
\begin{tabular}{rrrrrr}
  \hline
 Contaminated sites & 0 & ]1;5] & ]5;10] & ]10;20] & ]20;58]\\
  \hline
 Percent of UAP Ds (out of 367)& 50.7 & 31.9 & 7.0 & 6.7 & 3.8  \\
Percent of pixels (out of 9480) & 80.3 & 16.9 & 1.9 & 0.7 & 0.2 \\
   \hline
\end{tabular}\end{table}

For creating a pixel image of this variable, we have chosen as covariate the kernel smoothed intensity of the contaminated communes considered as a point pattern, including the number of contaminated sites as weight. The value of $\sigma$ has been taken equal to 5 kilometers, the same as for the population density covariate. Fig.~\ref{contaminated} on the right panel shows an example of computing of this variable. It is obvious that the covariate thus created is correlated to population density, but the spatial distribution of these two covariates are not exactly the same (see Fig.~\ref{covariates1}).
Finally, the Spatial Kolmogorov-Smirnov test of CSR also rejects the hypothesis of CSR with a $D$ test statistic equal to 0.3154 (associated to a p-value lower than 0.05), showing that the contaminated land covariateis significant to explain the intensity of the UAPs.

\subsection{Other covariates}
We also consider including the following variables:
\begin{itemize}
\item The percentage of wetlands and water bodies.
\item The kernel smoothed intensity of airport installations including runways, buildings and associated land.
\item The percentage of land occupied by forests.
\item The yearly sum of global irradiation on vertical surface (kWh/$m^2$) \citep{solar}.
\item The kernel smoothed intensity of the UAP As.
\end{itemize}
The first three variables originate from the Corine Land Cover France 2006 database (Source : European Union – SOeS, Corine Land Cover, 2006) and are given for each of the communes in France. We use these variables for the following reasons: The areas with a high rate of moisture could be more exposed to some aerial phenomenon such as a rise of moisture laden air. %(see for example: \url{http://www.cnes-jeunes.fr/web/CNES-Jeunes-fr/8166-des-faits-reels-saupoudres-de-fantasmagories.php}).
An aircraft can be interpreted as an unidentified aerial object under many circumstances, by both day and night. % (See for example \url{http://www.uapreporting.org/?page_id=346}).
Thus, it seems interesting to use the number of airport installations in the neighborhood. Several testimonies mentioned that the phenomenon was observed at the proximity of some forest. To compute the pixel images from these 3 variables, we use exactly the same method as previously.

The yearly sum of global irradiation on vertical surface has been included in our model because it could be that the areas where the sky is clear of clouds (hence with a strong value of global irradiation) make the UAP Ds more visible in the sky. This information was available at the pixel level but on a different grid (see \url{http://re.jrc.ec.europa.eu/pvgis/download/download.htm}). So, we transform the grid by a spatial smoothing as we did previously for the population density.

Finally, we also include the kernel smoothed intensity of the UAP As as covariate in the model. Indeed, if some cases totally explained by the GEIPAN are located in some concentrated areas, we can suppose that these areas also contains cases classified as UAP Ds. The value of $\sigma$ has been taken equal to 20 kilometers.

Each of the Kolmogorov-Smirnov test reject the hypothesis of CSR (see Table~\ref{KStest}) showing that the covariates are individually significant to explain the intensity of the UAP Ds. Finally, the covariates are represented on Fig.~\ref{covariates2}.

\begin{table}
\caption{\label{KStest} Kolmogorov-Smirnov test.}
\centering
\small{
\begin{tabular}{rrrrrr}
  \hline
 Covariates & Wetlands & Airport & Forest & Sun & UAP A\\
  \hline
 $D$ statistic & 0.1902 & 0.2569 & 0.089 & 0.1685 & 0.1782 \\
p-value & $2.155\times10^{-12}$ & $<10^{-16}$ &  0.004782 &  $8.083\times10^{-10}$ & $6.198\times10^{-11}$ \\
   \hline
\end{tabular}
}
\end{table}

\section{Model}\label{section4}
Using R, \citet{Baddeley} propose to fit the inhomogeneous Poisson model with the function \textit{ppm()}. It computes the intensity as a loglinear function of the covariates $\mathbf{Z}$:
\begin{center}
$\lambda(u) = e^{\mathbf{Z(u)}\beta }$
\end{center}
where $\beta$ are the parameters $[\begin{array}{c|c|c}
\beta_0 &\cdots & \beta_8
\end{array}]$ associated to the covariates $\mathbf{Z}$:
\begin{itemize}
\item $z_0$: Constant equal to 1,
\item $z_1$: Logarithm of the population density (\emph{pop}),
\item $z_2$: Kernel smoothed intensity of contaminated land (\emph{conta}),
\item $z_3$: Kernel smoothed intensity of the number of nuclear sites (\emph{nuclear}),
\item $z_4$: Percentage of wetlands (\emph{wetlands}),
\item $z_5$: Logarithm of the yearly sum of global irradiation on vertical surface (\emph{sun}),
\item $z_6$: Kernel smoothed intensity of the airport installations (\emph{airport}),
\item $z_7$: Percentage of land occupied by forests (\emph{forests}).
\item $z_8$: Kernel smoothed intensity of the UAP A (\emph{uapA}).
\end{itemize}

We apply the logarithm transformation to the covariates with a heavy tail distribution. The model may be fitted by the method of maximum pseudolikelihood by specifying as quadrature scheme the 381 UAPs as data points and the 9480 pixels as dummy points \citep{Baddeley00practicalmaximum}. The results are given in Table~\ref{result}.

\begin{table}
\caption{\label{result} Results of the modeling with all covariates.}
\centering
\begin{tabular}{rrrr}
  \hline
 & coefficients & std & p-value\\
  \hline
  (Intercept) &  -5.83328 & 3.81555 & 0.12634 \\
  log(pop)  & 0.49218 & 0.06187 & $<10^{-16}$ \\
  conta &   2.62161 & 0.91489 & 0.00417 \\
  nuclear & 256.77992 & 66.66179 & 0.00012 \\
  wetlands &  0.02691 & 0.01484 & 0.06987 \\
  log(sun)  & -0.53930 & 0.53930 & 0.31734 \\
  airport &  0.09101 & 0.06677 & 0.17287 \\
  forests & 0.00393 & 0.00274 & 0.15177 \\
  uapA &   64.80815 & 78.93748 & 0.41166 \\
   \hline
\end{tabular}\end{table}

We removed step by step the least significant variable. We removed first the covariate \emph{uapA} (p-value=0.41), then \emph{log(sun)} (p-value=0.35), then  \emph{forests} (p-value=0.22), \emph{aero} (p-value=0.11), and finally \emph{wetlands} (p-value=0.12). We finally choose to keep the model presented in Table~\ref{result2}.

\begin{table}
\caption{\label{result2} Results of the final modeling.}
\centering
\begin{tabular}{rrrr}
  \hline
 & coefficients & std & p-value \\
  \hline
(Intercept) &  -9.67973 & 0.21282  & $<10^{-16}$ \\
  log(pop) & 0.54115 & 0.04684  & $<10^{-16}$\\
  conta &  2.39716 & 0.86181 & 0.00542 \\
  nuclear & 247.07907 & 64.55783 & 0.00013 \\
   \hline
\end{tabular}\end{table}

To interpret correctly the values of the coefficients $\beta_j$, we need to calculate the derivative of the estimated function $\hat \lambda$ at pixel $u$, with respect to $z_j$:
\begin{eqnarray*}
 \frac{\partial}{\partial z_j}\hat \lambda(u) &=&  \frac{\partial}{\partial z_j}(exp(\displaystyle\sum_{i=0}^8z_i\hat\beta_i))\\
              &=& \hat\beta_j (exp(\displaystyle\sum_{i=0}^8z_i\hat\beta_i))
\end{eqnarray*}

We see that the partial derivative of $\hat \lambda$ is positive provided that $\hat\beta_j$ is positive: In this case, the higher the value of $z_j$, the higher the intensity. Thus, all the covariates in our model have a positive contribution on $\hat\lambda$ because the coefficients $\beta_j$ are positive.

We now consider the pixel $u^{*}$ taken as an example in this article. The estimated values for the different covariates are given in Table~\ref{zvalues}. Thus the estimated intensity $\hat\lambda$ at pixel $u^{*}$ is equal to:
\begin{eqnarray*}
\hat\lambda(u^{*}) &=&  e^{-9.679} \times 19.83 \times 1.09 \times  1.17 \\
              &\simeq& 0.00158\\
\end{eqnarray*}
\begin{table}
\caption{\label{zvalues} Values of the covariates estimated at the pixel $u^{*}$.}
\centering
\begin{tabular}{rrrr}
  \hline
 & pop  & conta & nuclear \\
  \hline
 $z_j(u^{*})$ & 5.521 & 0.036 & 0.00062 \\
  $z_j(u^{*})\times\beta_j$ & 2.99 & 0.087 & 0.15 \\
  $e^{z_j(u^{*})\times\beta_j}$ & 19.83 & 1.09 &  1.17   \\
   \hline
\end{tabular}\end{table}

\citet{citeulike:390763} define the raw residuals for spatial point processes as:
\begin{center}
$s(u)=\lambda^{*}(u)-\hat\lambda(u)$
\end{center}
In the particular case of $u^{*}$:
$s(u^{*})=\lambda^{*}(u^{*})-\hat\lambda(u^{*})\simeq0.00168-0.00158=1.05\times10^{-4}$.

\begin{figure}
\centering
\makebox{\includegraphics[width=8.5cm, trim=0cm 0cm 0cm 0cm, clip=true]{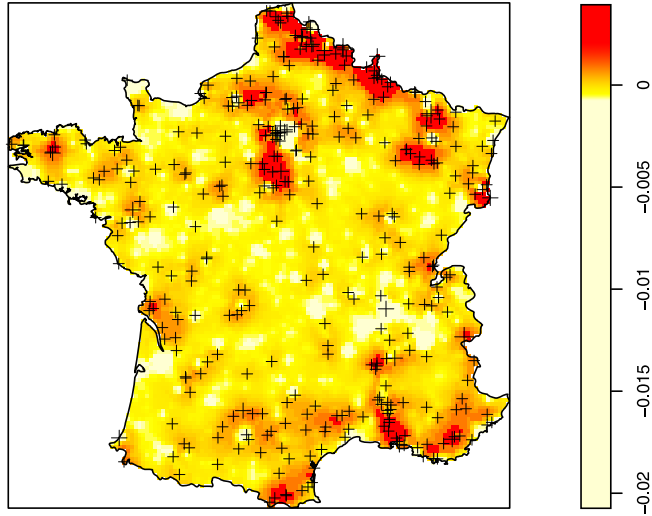}}
\caption{\label{residuals} Residuals of the model. The $5\%$ highest are shown in red. The $5\%$ lowest are shown in pale yellow.}
\end{figure}

The map of the Pearson residuals ($s(u)/\hat\lambda(u)$) is presented in Fig.~\ref{residuals}. The $5\%$ areas which have been the most underestimated by the model appear in red. It corresponds to the Belgian border, the tip of Britany, some parts in the South-East, the Picardie and Haute-Normandie regions, the Loiret and Corr\`eze departments. The $5\%$ areas which have been the most overestimated essentially correspond to the areas with high values of population density but without many UAP As at their proximity, such as the large French cities of Lyon, Bordeaux or Toulouse. The Paris region appears as an outlier because there are many UAP Ds in that area and finally the cluster of UAP Ds is bordered by an underestimated area (the city of Paris) and an overestimated area (the western suburbs of Paris with a lower population density).

\section{Conclusion}\label{section5}
\label{section5} When a phenomenon is graded D, it is extremely difficult for the GEIPAN to convincingly
establishits objective existence because the public's opinion on this hard-to-tackle subject is
often tainted with suspicion. Moreover, the public can always doubt that any testimony either be a hallucination or a hoax.
For this reason, the purpose of this study was not to bring a new explanation for the UAP Ds,
but to determine whether there is a link between the locations of the UAP Ds and some covariates provided by the different
French statistics services, and therefore open new, more precise and qualitative research avenues.

This study, conducted using the tools of the spatial point pattern analysis, reveals that,
the localization of the UAP Ds can indeed partly be explained by anthropogenic covariates.
The link between nuclear activities and UAP Ds, which has long been suspected and considered,
is now for the first time measured and appears surprisingly high (\textit{p}-value: 0.00013).
We also discovered a strong relationship between UAP Ds and contaminated land (\textit{p}-value:
0.00542) which until now had never been addressed.

These correlations can either be the result of an emerging endogenous activity,or of exogenous activity.
One open hypothesis is that these sensitive sites may be places of interest because of their connection with environmental issues.
However, we found in the analysis of the residuals that some clusters of UAP Ds are still not explained by the model.
A new track would be to include some covariates reflecting the level of education of the population.

Finally, GEIPAN might be more interested in these clusters by looking at the data on the sociology and psychology,
so information on witnesses and investigative practices.

\section*{Acknowledgements}
The authors would like to thank the team of students who participated in this study (Xusong
Bao, Maha Mazouzi, Tony Ong, Cecilia Rivera-Martinez) and the French Space Agency CNES/GEIPAN for making these data available. This work was supported by the Agence Nationale de la Recherche under
the ModULand project (ANR-11-BSH1-005).
\newpage
\bibliographystyle{chicago}
\bibliography{bibliototal}

\appendix
\section{Descriptive statistics}

\begin{table}
\caption{\label{descriptable}Descriptive statistics of the covariates computed on the 9480 pixels included in $A$.}
\centering
\begin{tabular}{rrrrrrr}
  \hline
 & Min. & 1st Qu. & Median & Mean & 3rd Qu. & Max. \\
  \hline
population & 2.18 & 24.66 & 45.28 & 102.80 & 89.12 & 10380 \\
  contaminated land & 0 & 0 & 0 & 0.01 & 0.01 & 0.67 \\
  nuclear sites & 0 & 0 & 0 & 0.20 & 0 & 13 \\
  wetlands & 0 & 0.03 & 0.22 & 0.93 & 0.87 & 80.48 \\
  sun & 998 & 1144 & 1227 & 1241 & 1323 & 1581 \\
  airport & 0 & 0 & 0 & 0.08 & 0.01 & 14.59 \\
  forest & 0.09 & 11.40 & 24.28 & 31.48 & 45.94 & 99.25 \\
   \hline
\end{tabular}
\end{table}

\section{Mapping of the covariates}\label{appendix1}

\begin{figure}
\centering
\makebox{\includegraphics[trim=0cm 1cm 0cm 0.2cm, clip=true]{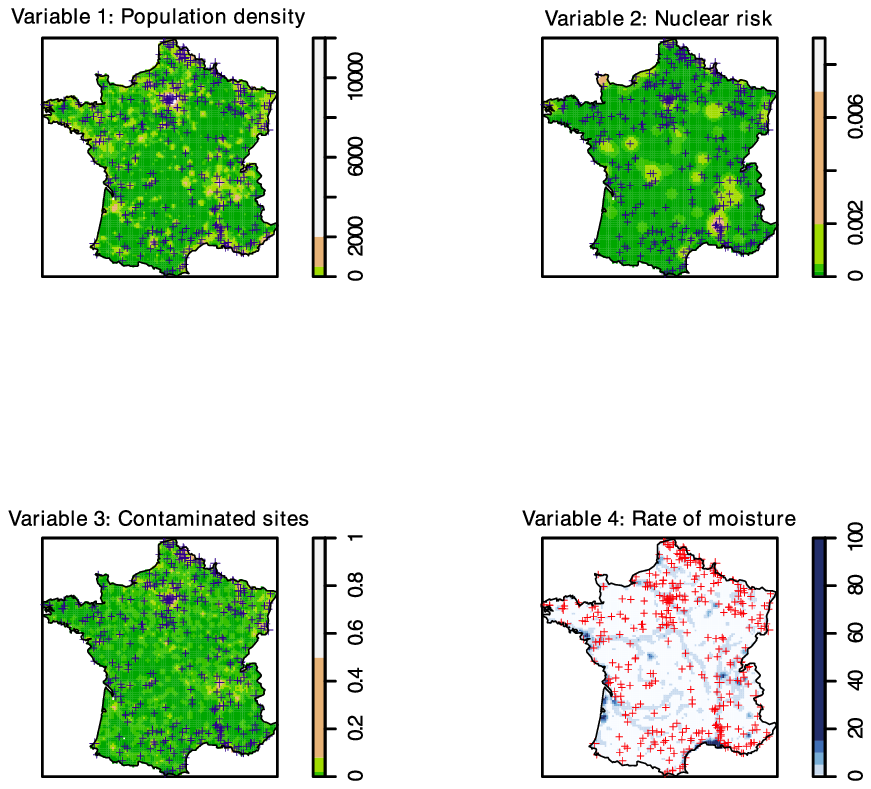}}
\caption{\label{covariates1} Representation of the covariates transformed as pixel images (1).}
\end{figure}

\begin{figure}
\centering
\makebox{\includegraphics[trim=0cm 1cm 0cm 0.2cm, clip=true]{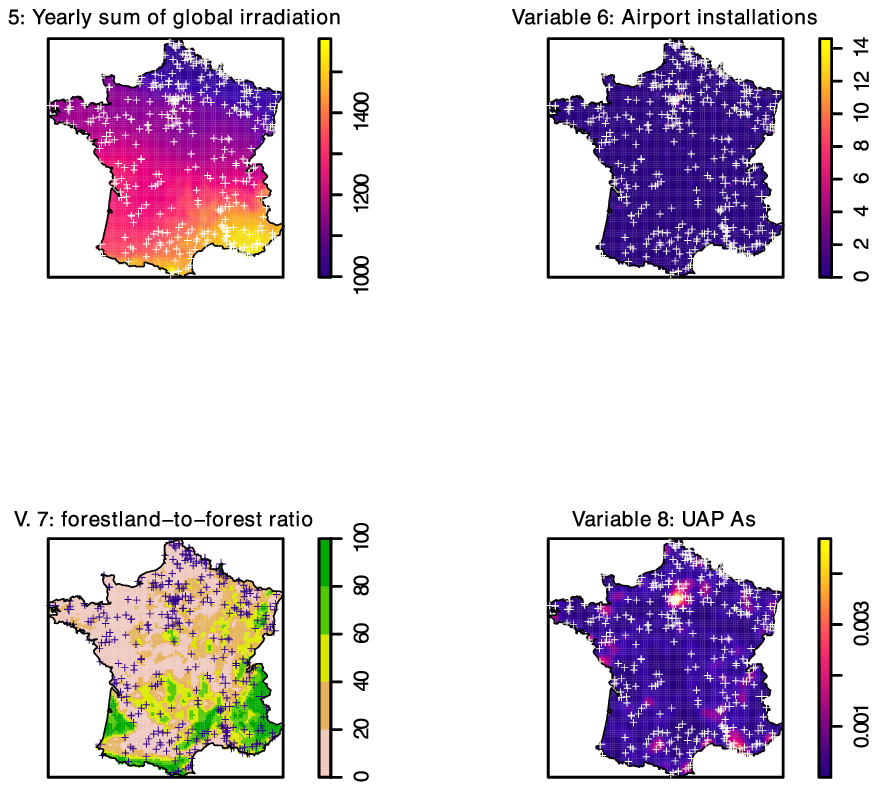}}
\caption{\label{covariates2} Representation of the covariates transformed as pixel images (2).}
\end{figure}

\end{document}